\documentclass[aps,amsmath,amssymb,twocolumn]{revtex4-2}

\UseRawInputEncoding
\usepackage{amsmath} 
\usepackage{amssymb} 
\usepackage{graphicx}
\usepackage{dcolumn}
\usepackage{bm}
\usepackage{color}
\usepackage[normalem]{ulem}

\begin{document}

\title{Deep learning for the modeling and inverse design of radiative heat transfer}

\author{J.~J. Garc\'{\i}a-Esteban}
\author{J. Bravo-Abad}
\author{J.~C. Cuevas}

\affiliation{Departamento de F\'{\i}sica Te\'orica de la Materia Condensada
and Condensed Matter Physics Center (IFIMAC), Universidad Aut\'onoma de Madrid,
E-28049 Madrid, Spain}

\date{\today}

\begin{abstract}
Deep learning is having a tremendous impact in many areas of computer science and engineering. Motivated 
by this success, deep neural networks are attracting an increasing attention in many other disciplines, 
including physical sciences. In this work, we show that artificial neural networks can be successfully used 
in the theoretical modeling and analysis of a variety of radiative heat transfer phenomena and devices. By 
using a set of custom-designed numerical methods able to efficiently generate the required training datasets, 
we demonstrate this approach in the context of three very different problems, namely, \emph{(i)} near-field 
radiative heat transfer between  multilayer systems that form hyperbolic metamaterials,  \emph{(ii)} passive 
radiate cooling in photonic-crystal slab structures, and \emph{(iii)} thermal emission of subwavelength objects. 
Despite their fundamental differences in nature, in all three cases we show that simple neural network architectures 
trained with datasets of moderate size can be used as fast and accurate surrogates for doing numerical simulations, 
as well as engines for solving inverse design and optimization in the context of radiative heat transfer. Overall, 
our work shows that deep learning and artificial neural networks provide a valuable and versatile toolkit for 
advancing the field of thermal radiation.
\end{abstract}

\maketitle

\section{Introduction} \label{sec-intro}

Deep learning is a form of machine learning that allows a computational model composed of multiple layers
of processing units (or artificial neurons) to learn multiple levels of abstraction in given data 
\cite{LeCun2015,Goodfellow2016,Aggarwal2018}. In recent years, there has been a revival of deep learning
triggered by the availability of large datasets and recent advances in architectures, algorithms, and computational 
hardware \cite{LeCun2015}. This, in turn, has resulted in a huge impact of deep learning in topics related to computer 
science and engineering such as computer vision \cite{Krizhevsky2012}, natural language processing \cite{Cho2014}, 
autonomous driving \cite{Shalev-Shwartz2016}, or speech recognition \cite{Hinton2012}, just to mention a few. Motivated 
by this success, deep learning is attracting an increasing attention from researchers in other disciplines. In 
particular, deep learning and artificial neural networks have already found numerous applications in the physical 
sciences, see recent reviews of Refs.~\cite{Mehta2019,Carleo2019,Marquardt2021}. 

A paradigmatic example is the field of photonics (including nanophotonics, plasmonics, metamaterials, etc.) in which all 
the basic types of neural networks have already been employed to model, design, and optimize photonic devices. The first 
applications of neural networks in photonics actually date back to the 1990s and were related to the computer-aided design 
of microwave devices \cite{Zhang2003}. But it has been only in the last three years that we have witnessed a true
revolution in this topic, for detailed reviews see Refs.~\cite{So2020,Hegde2020,Jiang2020,Ma2021,Piccinotti2021,Liu2021}. 
Neural networks in photonics are being used for three main purposes. First, deep neural networks, configured as 
discriminative networks, are used to do forward modeling of photonic structures, i.e., they are operated as high-speed 
surrogate electromagnetic solvers, see for instance Refs.~\cite{Peurifoy2018,Qu2019}. Second, it has been shown that properly 
trained networks can be efficiently used to optimize structures for a given purpose \cite{Peurifoy2018,Ma2018,Jiang2019a}. 
Third, neural networks are used to tackle inverse design problems \cite{Liu2018,An2019,Jiang2019b,Unni2020}, and, once trained, 
they have been shown to be clearly faster for this task than other existent numerical strategies \cite{Peurifoy2018}.

Radiative heat transfer \cite{Modest2013,Howell2016,Zhang2007} is experiencing its own revival in recent years 
\cite{Cuevas2018}. Thus, for instance, the study of the thermal radiation exchange in the near-field regime is attracting a lot 
of attention \cite{Basu2009,Song2015,Cuevas2018,Biehs2021}. The progress on this topic includes crucial experimental advances and 
numerous theoretical proposals to tune, actively control, and manage near-field thermal radiation. Other topics of great current 
interest in this field are the control of thermal emission of an object, with special emphasis in its implications 
for energy applications \cite{Li2018,Fan2019}, and the comprehension of far-field radiative heat transfer beyond Planck's law 
\cite{Cuevas2019}. Despite its significant potential, and apart from notable exceptions \cite{Kudyshev2020}, a systematic study 
of the application of deep learning techniques to radiative heat transfer is still lacking.

In this work we show how artificial neural networks can be helpful in the modeling and analysis of a wide variety of
thermal radiation phenomena, as well as in the optimization and inverse design of structures for radiative heat transfer.   
To illustrate these ideas we present here the use of neural networks in three distinct problems that cover many
of the basic aspects of current interest in the field of radiative heat transfer. In the first example, we use neural 
networks in the context of near-field radiative heat transfer between multilayer systems that form hyperbolic metamaterials. 
In a second example, we show how neural networks can be used to optimize the performance of a device in the context of 
passive radiative cooling. In the third example, we illustrate how neural networks can be helpful in the description of 
the thermal emission of an object of arbitrary size and shape, with especial attention to subwavelength objects. In all 
three cases we used custom-designed numerical methods that allow us to carry out an efficient and robust generation of 
the training data required in the proposed approach.

The rest of the paper is organized as follows. In Sec.~\ref{sec-NN} we briefly introduce the topic of neural networks,
as used in this work, and discuss some of the technical details related to their practical implementation. In 
Sec.~\ref{sec-NFRHT} we discuss the use of neural networks in the context of the near-field radiative heat transfer 
between multilayer structures. Then, Sec.~\ref{sec-cooling} is devoted to the use of neural networks for the optimization 
of devices in the context of passive radiative cooling. In Sec.~\ref{sec-emission} we show how neural networks can be 
helpful in the problem of the description of the thermal emission of a single object of arbitrary size and shape. Finally, 
we summarize our main conclusions in Sec.~\ref{sec-conclusions}.

\section{Theoretical background} \label{sec-NN}

\begin{figure}[t]
\includegraphics[width=\columnwidth,clip]{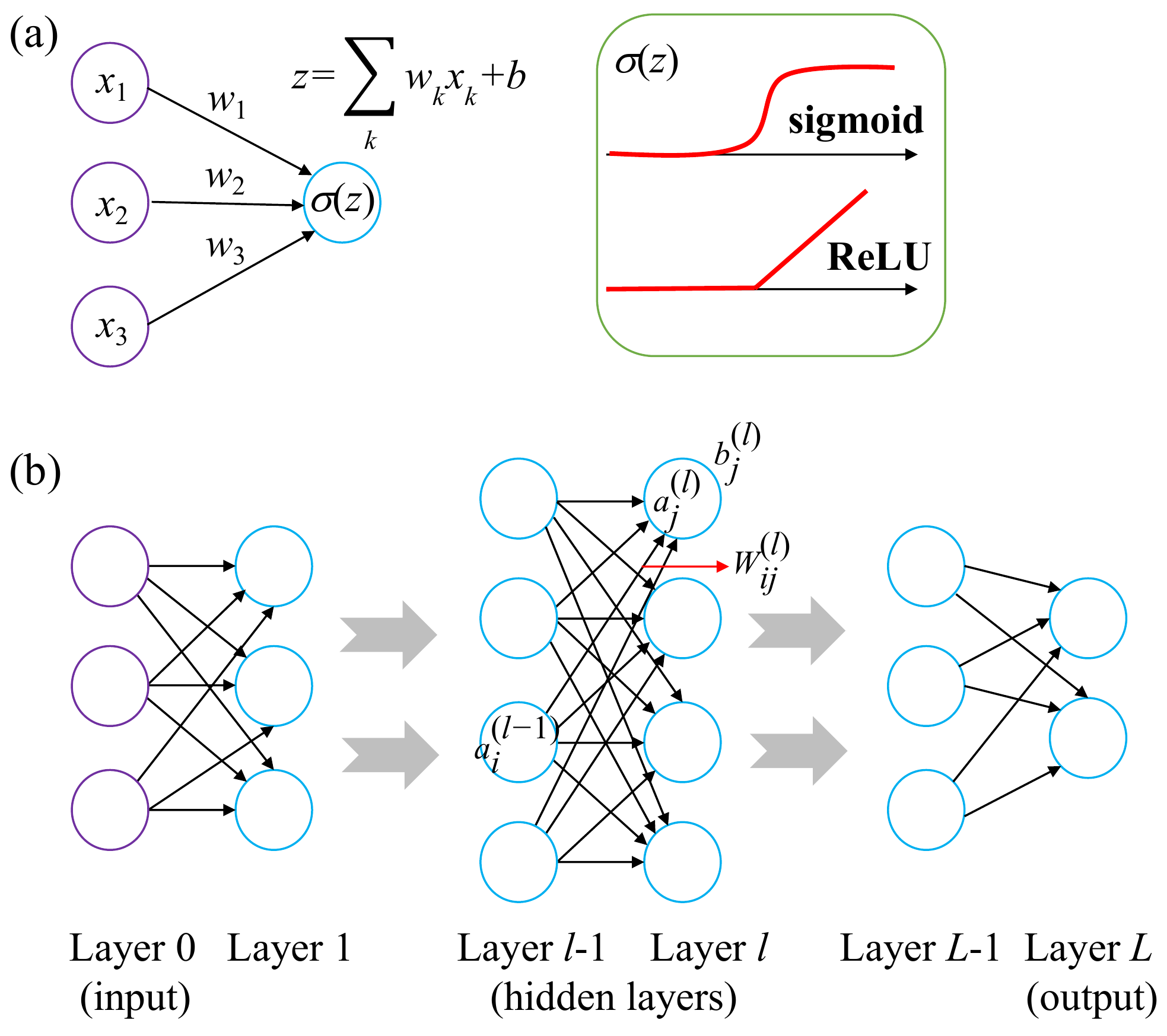}
\caption{(a) Schematic representation of a neuron, the basic component of a neural network. The value of a
neuron is determined by a linear transformation that weights the importance of various inputs, followed by a 
nonlinear activation function. Two typical nonlinear activation functions are also shown: sigmoid and ReLU.
(b) Feed-forward neural network with neurons arranged into layers with the output of one layer serving as the 
input to the next layer.}
\label{fig-NN}
\end{figure}

Neural networks (NNs), as used in this work, are nonlinear models for supervised learning. More specifically, 
they are general-purpose function approximators that can be trained using many examples. The basic unit of a 
NN is an artificial neuron that takes $n$ input features $\{x_1, x_2, \dots, x_n\}$ and produces a scalar
output $a(x_i)$, see Fig.~\ref{fig-NN}(a). The value of the neuron $a$ is obtained starting from the values
$x_k$ of some other neurons that feed into it as follows. First, one calculates a linear function of those 
values: $z = \sum_k w_k x_k + b$, where the coefficients $w_k$ are called the \emph{weights} and the offset $b$
is called the \emph{bias}. Then, a nonlinear function $\sigma$, known as \emph{activation function}, is applied to 
yield the neuron's value: $a = \sigma(z)$. There are many different choices for the activation function $\sigma$ 
and in this work we have mainly used two of the most popular ones, namely the sigmoid $\sigma(z) = 1/
(1+e^{-z})$ and the rectified linear unit (ReLU) $\sigma(z) = \mbox{max}(0,z)$, see Fig.~\ref{fig-NN}(a). 

A NN consists of many neurons stacked into layers, with the output of one layer serving as the input of the next one, 
see Fig.~\ref{fig-NN}(b). In this simple feedforward fully-connected network, which is the architecture used
throughout this work, the first layer ($l=0$) is called the \emph{input layer}, the middle layers (from $l=1$
to $l=L-1$) are called \emph{hidden layers}, and the final layer ($l=L$) is called the \emph{output layer}. Here, 
we have $n_l$ neurons in layer $l$. The task of this network is for every input to produce an output, which will 
depend on the current value of the parameters of the model (weights and biases). To see how the network operates,
let us consider a single input example with $n$ features encoded in the row vector $x = (x_1, x_2, \dots, x_n)$.
Let us call $W^{(l)}$ the $n_{l-1} \times n_{l}$ matrix whose element $W^{(l)}_{ij}$ is the weight connecting the neuron 
$i$ of layer $l-1$ with neuron $j$ of layer $l$, and $b^{(l)}$ the row vector whose element $b^{(l)}_j$ is the bias
corresponding to neuron $j$ in layer $l$. The output of the network is obtained by going layer by layer, starting at 
the input layer $l=0$, whose neuron value is simply the example $x$ provided by the user, i.e.,
\begin{eqnarray}
z^{(0)} & = & x , \;\; a^{(0)} = z^{(0)}  \nonumber \\
z^{(1)} & = & a^{(0)} W^{(1)} + b^{(1)} , \;\; a^{(1)} = \sigma(z^{(1)}) \nonumber \\
& \vdots & \nonumber \\
z^{(l)} & = & a^{(l-1)} W^{(l)} + b^{(l)} , \;\; a^{(l)} = \sigma(z^{(l)}) \nonumber \\
& \vdots & \nonumber \\
z^{(L)} & = & a^{(L-1)} W^{(L)} + b^{(L)} , \;\; \hat y = a^{(L)} = \sigma(z^{(L)})  .
\end{eqnarray}
Here, $z^{(l)}$ and $a^{(l)}$ are row vectors with $n_l$ elements and $\hat y$ is a row vector with
$n_L$ elements containing the output of the network corresponding to the input $x$. Moreover, 
$\sigma(z^{(l)})$ should be understood as the element-wise application of the activation function $\sigma$,
which could be different in different layers. These equations can be trivially generalized to deal in 
parallel with an arbitrary number of input examples.

Once a network architecture is fixed, the next step is to train the NN, i.e., to adjust the parameters of the
model (weights and biases) to reproduce the desired function. Like in all supervised learning 
procedures, one starts by providing a set of examples (or \emph{training set}) $\{(x_i, y_i); i=1, \dots, m\}$,
where $x_i$ contains the $n$ input variables or features of example $i$, and $y_i$ corresponds to the \emph{target}
variable or output of example $i$. In the problems addressed in the following sections, the features will be the 
geometrical parameters defining the investigated structures (thickness, size, filling factor, etc.) and the
number of these input variables will set the number of neurons of the input layer ($n_0=n$). On the other hand,
the target will correspond to a spectral function (like a spectral thermal conductance or a frequency-dependent 
emissivity) that will be calculated numerically using different computational methods. The number of neurons of 
the output layer $n_L$ will be equal to the number of frequency or wavelength points used to represent that spectral 
function, and the value of the output neurons will correspond to the prediction made by the NN for that function. 
To train the network one also needs an \emph{error function}, also known as \emph{cost or loss function}, that provides 
a metric of the deviation between the NN output and the function that it is trying to approximate. A typical choice, 
which we will frequently use, is the mean square error (MSE) given by
\begin{equation}
E(\theta) = \frac{1}{m} \sum^m_{i=1} (y_i - \hat y_i(\boldsymbol{\theta}))^2 ,	
\end{equation}
where $\boldsymbol{\theta}$ represents the set of model parameters (weights and biases), $m$ is the number of 
examples in the training set, $y_i$ is target $i$ and $\hat y_i$ is the NN prediction for example $i$. 

Once the cost function has been defined, the idea is to find its minimum in the high-dimensional space of 
parameters $\boldsymbol{\theta}$. This minimization can be done with the method of gradient descent, or any 
of its generalizations \cite{Mehta2019}, and a proper choice of the \emph{learning rate} (the parameter that 
determines how big a step we should take in the direction of the gradient). In this work we have always made 
use of the ADAM optimizer \cite{Kingma2014}, which has been shown to be a robust choice for deep learning 
optimization in a variety of different contexts \cite{Schmidt2021}. ADAM makes use of running averages of both 
the gradients of the cost function and their second moments. In general, we have used ADAM as a stochastic 
algorithm. Stochasticity is incorporated by approximating the gradient of the cost function on a subset 
of the data called a \emph{minibatch}, which has a size ($m_{\rm batch}$) much smaller than the number of training 
examples $m$. In every optimization step we use the mini-batch approximation to the gradient to update the parameters 
$\boldsymbol{\theta}$ and then we cycle over all $m/m_{\rm batch}$ minibatches one at a time. A full iteration over 
all $m$ data points (i.e., using all $m/m_{\rm batch}$ minibatches) is called an \emph{epoch}. 

Due to the large number of parameters $\boldsymbol{\theta}$, the training procedure of a NN requires a specialized 
algorithm, which is referred to as \emph{backpropagation} \cite{Rumelhart1985,Rumelhart1986}. This algorithm is 
conceptually very simple and makes a clever use of the chain rule for derivatives of a multivariate function to compute 
the gradient of the cost function with respect to all the parameters in only one backward pass from the output layer
to the input layer. This algorithm is described in many textbooks and reviews, see e.g.\ 
Refs.~\cite{Rumelhart1986,Mehta2019,Marquardt2021}, and we have made use of it in all our calculations.

Another point worth mentioning is that, following the common practice in supervised learning, and in order to 
evaluate the ability of our models to generalize to previously not seen data, we have divided our training sets 
into two portions, the dataset we train on, which we shall simply refer to as training set, and a smaller 
\emph{validation} (or \emph{cross-validation}) \emph{set} that allows us to gauge the out-of-sample performance 
of the model. While training our models, we have followed both the training error and the validation error (using the 
same cost function). Then, we have adjusted the \emph{hyperparameters}, such as the number of layers and neurons
per layer, to reduce the validation error to optimize the performance for a specific dataset. Additionally, one 
should also use a third and independent portion of the original dataset as \emph{test set}, i.e., as a set that is 
neither used for training nor for validation and that provides an unbiased measure of the generalization ability of a 
model. This test set is strictly necessary when one uses any regularization method based on the validation error such 
as the so-called \emph{early stopping}, otherwise the validation set plays very much the role of the test set and
this latter one becomes unnecessary. Finally, we point out that most of our NN calculations were done using the 
open source library \emph{Tensorflow} \cite{TensorFlow}, and in particular, its higher level application programming 
interface (API) \emph{Keras} \cite{Keras}.

\section{Near-field radiative heat transfer between multilayer systems} \label{sec-NFRHT}
One of the major advances in recent years in the field of thermal radiation has been the experimental
confirmation of the long-standing prediction that the limit set by Stefan-Boltzmann's law for the radiative 
heat transfer between two bodies can be largely overcome by bringing them sufficiently close \cite{Polder1971}.
This phenomenon is possible because in the near-field regime, i.e., when the separation between two bodies 
is smaller than the thermal wavelength $\lambda_{\rm Th}$ ($\sim$10 $\mu$m at room temperature), radiative
heat can also be transferred via evanescent waves (or photon tunneling). This mechanism provides an additional 
contribution not taken into account in Stefan-Boltzmann's law and it turns out to dominate the near-field 
radiative heat transfer (NFRHT) for sufficiently small gaps o separations. Up to date, this phenomenon has been confirmed 
in numerous experiments and it has led to a huge experimental and theoretical activity on the topic of NFRHT, 
for recent reviews see Refs.~\cite{Song2015,Cuevas2018,Biehs2021}. The workhorse geometry in the study of NFRHT is 
that of two parallel plates and to maximize the heat transfer special attention has been devoted to the use of 
materials that exhibit electromagnetic surface modes at the body-vacuum interfaces, such as polar dielectrics 
(SiO$_2$, SiN, etc.) or metallic materials exhibiting surface plasmon polaritons in the infrared. Different strategies
have been recently proposed to further enhance NFRHT \cite{Song2015,Cuevas2018,Biehs2021}. One of the most popular
ideas is based on the use of multiple surface modes that can naturally appear in multilayer structures. In this
regard, a lot of attention has been devoted to multilayer systems where dielectric and metallic layers are
alternated to give rise to hyperbolic metamaterials \cite{Guo2012,Biehs2012,Guo2013,Biehs2013,Bright2014,
Miller2014,Biehs2017,Iizuka2018,Song2020,Moncada-Villa2021}. The hybridization of surface modes appearing
in different metal-dielectric interfaces have indeed been shown to lead to a great enhancement of the NFRHT,
as compared to the case of two infinite parallel plates, see e.g.\ Ref.~\cite{Iizuka2018}. The goal of this 
Section is to show how NNs can be used to describe the NFRHT between multilayer structures and how they can
assist in the design and optimization of these structures for different purposes. 

\begin{figure}[t]
\includegraphics[width=0.95\columnwidth,clip]{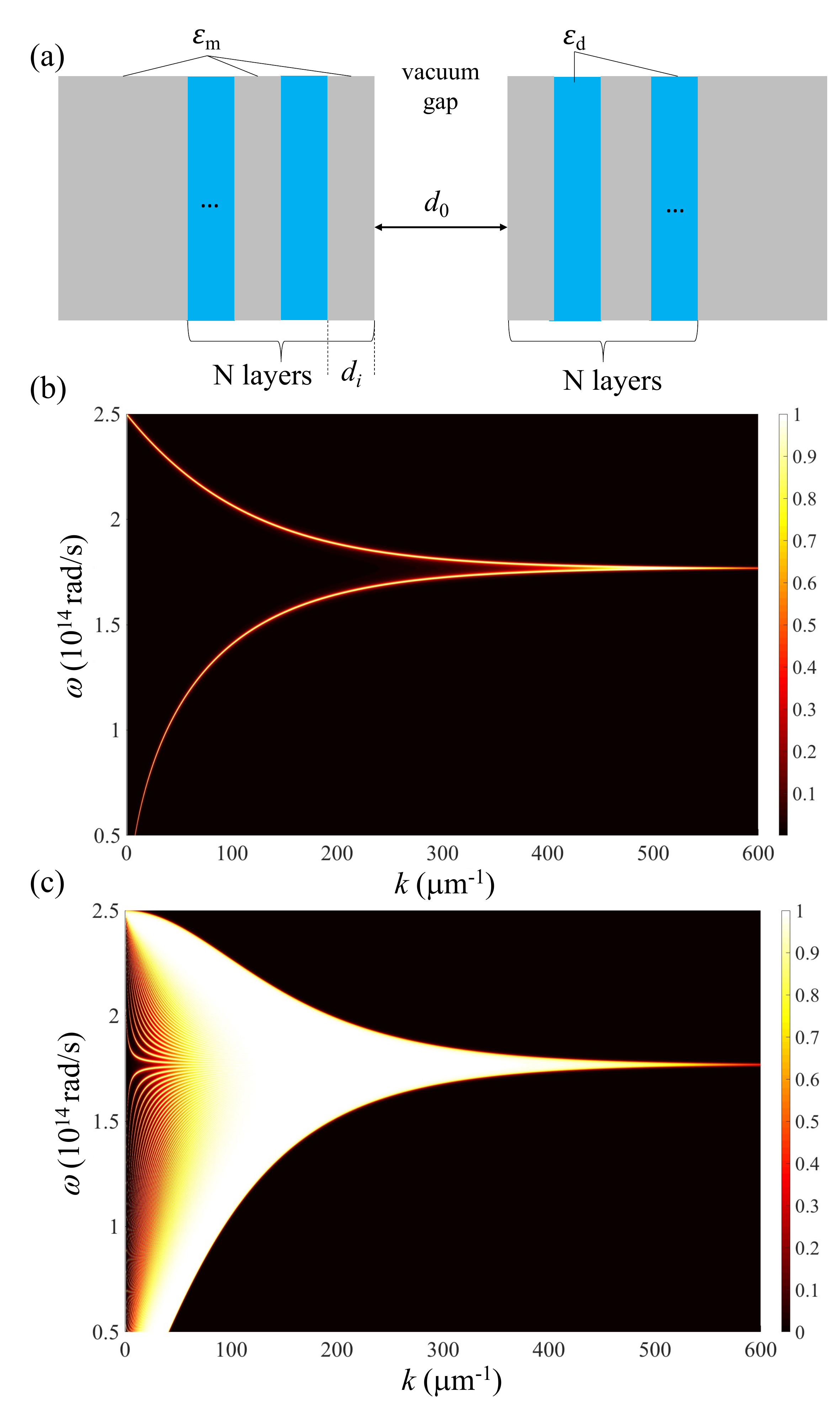}
\caption{(a) Sketch of two identical multilayer systems separated by a vacuum gap of size $d_0$. The two reservoirs 
feature $N$ total layers alternating between a Drude metal (grey areas) with permittivity $\epsilon_\mathrm{m}$ 
and a dielectric (white areas) with permittivity $\epsilon_\mathrm{d}$. The last layer in both cases is made 
of metal and the thickness of layer $i$ is denoted by $d_i$. (b) The transmission of the evanescent waves as 
a function of the frequency $\omega$ and the magnitude of the parallel wave vector $k$ for the bulk system, i.e., 
two parallel plates made of the metal, and $d_0 = 10$ nm. (c) The same as in panel (b), but for the multilayer 
system with $N= 160$ and $d_i = 10$ nm for all layers.}
\label{fig-multilayer1}
\end{figure}

Following Ref.~\cite{Iizuka2018}, we consider here the radiative heat transfer between two identical
multilayer structures separated by a gap $d_0$, as shown in Fig.~\ref{fig-multilayer1}(a). Each body contains 
$N$ total layers alternating between a metallic layer with a permittivity $\epsilon_\mathrm{m}$ and a lossless 
dielectric layer of permittivity $\epsilon_\mathrm{d}$. The thickness of the layer $i$ is denoted by $d_i$ and 
it can take any value within a given range (to be specified below). While the dielectric layers will be set to 
vacuum ($\epsilon_\mathrm{d} =1$), the metallic layers will be described by a permittivity given by a Drude model:
\begin{equation}
\epsilon_\mathrm{m}(\omega) = \epsilon_{\infty} - \frac{\omega^2_p}{\omega (\omega + i \gamma)} ,	
\end{equation}
where $\epsilon_{\infty}$ is the permittivity at infinite frequency, $\omega_p$ is the plasma frequency,
and $\gamma$ de damping rate. From now on, we set $\epsilon_{\infty} = 1$, $\omega_p = 2.5 \times 10^{14}$
rad/s, and $\gamma = 1 \times 10^{12}$ rad/s. 

We describe the radiative heat transfer within the framework
of theory of fluctuational electrodynamics \cite{Rytov1953,Rytov1989} and focus on the near-field regime.
In this regime, the radiative heat transfer is dominated by TM- or $p$-polarized evanescent waves and the heat 
transfer coefficient (HTC) between the two bodies, i.e., the linear radiative thermal conductance per unit of 
area, is given by \cite{Basu2009}
\begin{equation}
h = \frac{\partial}{\partial T} \int^{\infty}_0 \frac{d\omega}{2\pi} \Theta(\omega, T) 
\int^{\infty}_{\omega/c} \frac{dk}{2\pi} \, k \tau_p(\omega, k) ,
\end{equation}
where $T$ is temperature, $\Theta(\omega, T)= \hbar \omega/ (e^{\hbar \omega/ k_{\rm B} T} -1)$ is the mean 
thermal energy of a mode of frequency $\omega$, $k$ is the magnitude of the wave vector parallel to the surface 
planes, and $\tau_p(\omega, k)$ is the transmission (between 0 and 1) of the $p$-polarized evanescent modes given by
\begin{equation}
 \tau_p(\omega, k) = \frac{4 \left[ \mbox{Im} \left\{ r_p(\omega,k) \right\} \right]^2 e^{-2q_0 d_0}}
 {| 1 - r_p(\omega,k)^2 e^{-2 q_0 d_0} |^2} .
\end{equation}
Here, $r_p(\omega,k)$ is the Fresnel reflection coefficient of the $p$-polarized evanescent waves from 
the vacuum to one of the bodies and $q_0 = \sqrt{k^2 - \omega^2/c^2}$ ($\omega/c < k$) is the wave number 
component normal to the layers in vacuum. The Fresnel coefficient needs to be computed numerically
and we have done it by using the scattering matrix method described in Ref.~\cite{Caballero2012}. 
In our numerical calculations of the HTC we also took into account the contribution of $s$-polarized modes, 
but it turns out to be negligible for the gap sizes explored in this work.  

Let us briefly recall that, as explained in Ref.~\cite{Iizuka2018}, the interest in the NFRHT in this multilayer
structures resides in the fact that the heat exchange in this regime is dominated by surfaces modes that can be
shaped by playing with the layer thicknesses. In the case of two parallel plates made of a Drude metal, the 
NFRHT is dominated by the two cavity surface modes resulting from the hybridization of the surface plasmon polaritons
(SPPs) of the two metal-vacuum interfaces \cite{Iizuka2018}. As shown in Fig.~\ref{fig-multilayer1}(b), these
two cavity modes give rise to two near-unity lines in the transmission function $\tau_p(\omega, k)$. Upon introducing
more internal layers with appropriate thickness, one can have NFRHT contributions from surface states at multiple 
surfaces, as we illustrate in Fig.~\ref{fig-multilayer1}(c) for the case of $N=160$ layers. As shown in 
Ref.~\cite{Iizuka2018}, the contribution of these additional surface states originating from internal layers can 
lead to a great enhancement of the NFRHT as compared to the bulk system (two parallel plates) in a wide range of 
gap values. Our goal in this Section is to show that NNs can learn the NFRHT characteristics of these multilayer 
systems and that they can be used in turn to solve inverse design and optimization problems in this context.

\begin{figure}[t]
\includegraphics[width=\columnwidth,clip]{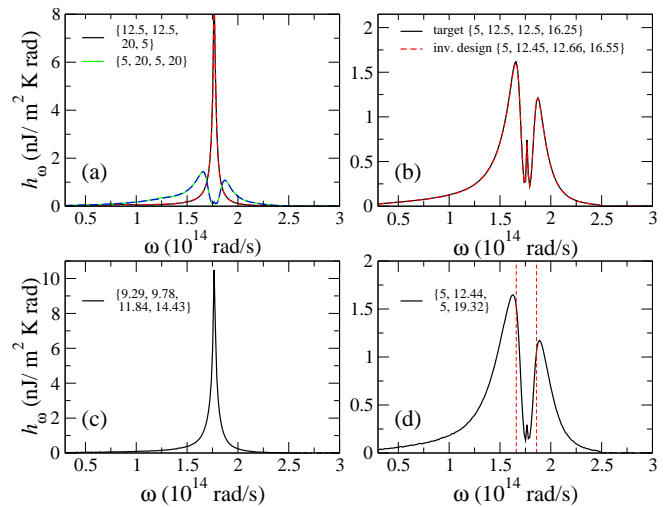}
\caption{ Results for the spectral heat transfer coefficient $h_\omega$ between two multilayers with $N=4$ and a gap 
size $d_0 = 10$ nm. (a) Comparison between real $h_\omega$-spectra computed with fluctuational electrodynamics (solid 
lines) and the prediction of the NN (dashed lines). The layer thicknesses (in nm) are indicated in the legend. 
(b) Comparison of the NN approximation to the target spectrum (layer thicknesses in nm are indicated in the legend) 
following the inverse design problem described in the text. (c) Result of the optimization problem in which the total
heat transfer coefficient is maximized. (d) Result of the optimization problem where $h_\omega$ is minimized in
the frequency region indicated by the dashed vertical lines.}
\label{fig-multilayer2}
\end{figure}

We start by considering the NFRHT between two systems formed by $N=4$ layer (2 metallic and 2 dielectric layers).
We set the vacuum gap to $d_0 = 10$ nm and the temperature to $T=300$ K. Our objective is to show that a NN can
learn, in particular, the spectral HTC $h_\omega$, i.e., the HTC per unit of frequency: $h = \int^{\infty}_0 h_\omega 
d\omega$. To train the network, we used the theory detailed above and prepared a training set with 881 
$h_\omega$-spectra where the thicknesses $d_i$ of the 4 layers were varied between 5 and 20 nm. Every spectrum
contains 200 frequency points in the range $\omega \in [0.3, 3] \times 10^{14}$ rad/s. The training set was, in turn,
divided into 80\% for actual training and 20\% for the test set. In this case we found that a NN with 5 hidden layers 
and 250 neurons per layer was able to accurately reproduce the training set. This network contains 4 neurons in the 
input layer (corresponding to the 4 input parameters, i.e., the layer thicknesses), while the output layer has 200 
neurons corresponding to the frequency values in the $h_\omega$-spectra. The NN was trained during 50,000 epochs using 
the MSE as the cost function, the ADAM optimizer, the ReLU activation function in all layers, except 
in the output one, and we did not use early stopping. We found helpful to use a variable learning rate
($l_r$) to improve the training given by $l_r = 0.001 \times 0.3^{p/20,000}$, where $p$ is the number of epochs (the 
numerical values defining $l_r$ have been found by a trial-and-error process). To give a quantitative idea about the 
ability of our network to reproduce the training set and to generalize, we calculated the average relative error in 
the integral of the $h_\omega$-spectra (i.e., the total HTC) and found that is 0.81\% for the training set and 1.45\% 
for the test set. The generalization accuracy of the NN is illustrated in Fig.~\ref{fig-multilayer2}(a) where we show 
how the network is able to reproduce different spectra from the test set (i.e., spectra it was not trained on).

As a next step, we show how the NN can be used to solve inverse design problems. As proof-of-principle calculation, the 
idea is to show that with the help of the NN we can find the layer thicknesses that would be able to reproduce an arbitrary 
$h_\omega$-spectrum. For this purpose, we freeze all the parameters of the NN and use backpropagation to train the inputs. 
This is done by fixing the output to the desired output and iterating the input to minimize the difference between the 
spectrum predicted by the NN and the target spectrum. This means in practice that the cost function for this task is simply 
defined as the MSE between the predicted and the target spectrum. This minimization process is very efficient because 
the gradients of the cost function with respect to the inputs can be computed analytically using backpropagation 
\cite{Peurifoy2018}. Once the minimization process is finished, the NN suggests the thickness values to reproduce the 
target spectrum. We illustrate this inverse design problem in Fig.~\ref{fig-multilayer2}(b) where the target spectrum was 
randomly chosen to correspond to the layer thicknesses $\{d_1, d_2, d_3, d_4\} = \{5.0, 12.5, 12.5, 16.25\}$ nm (to ensure 
that we have a physically realizable spectrum). Notice that the NN is able to reproduce this target spectrum very well and 
suggests that the corresponding layer thicknesses are $\{5.0, 12.45, 12.66, 16.55\}$ nm, which is in excellent agreement 
with the actual value.

\begin{figure*}[t]
\includegraphics[width=\textwidth,clip]{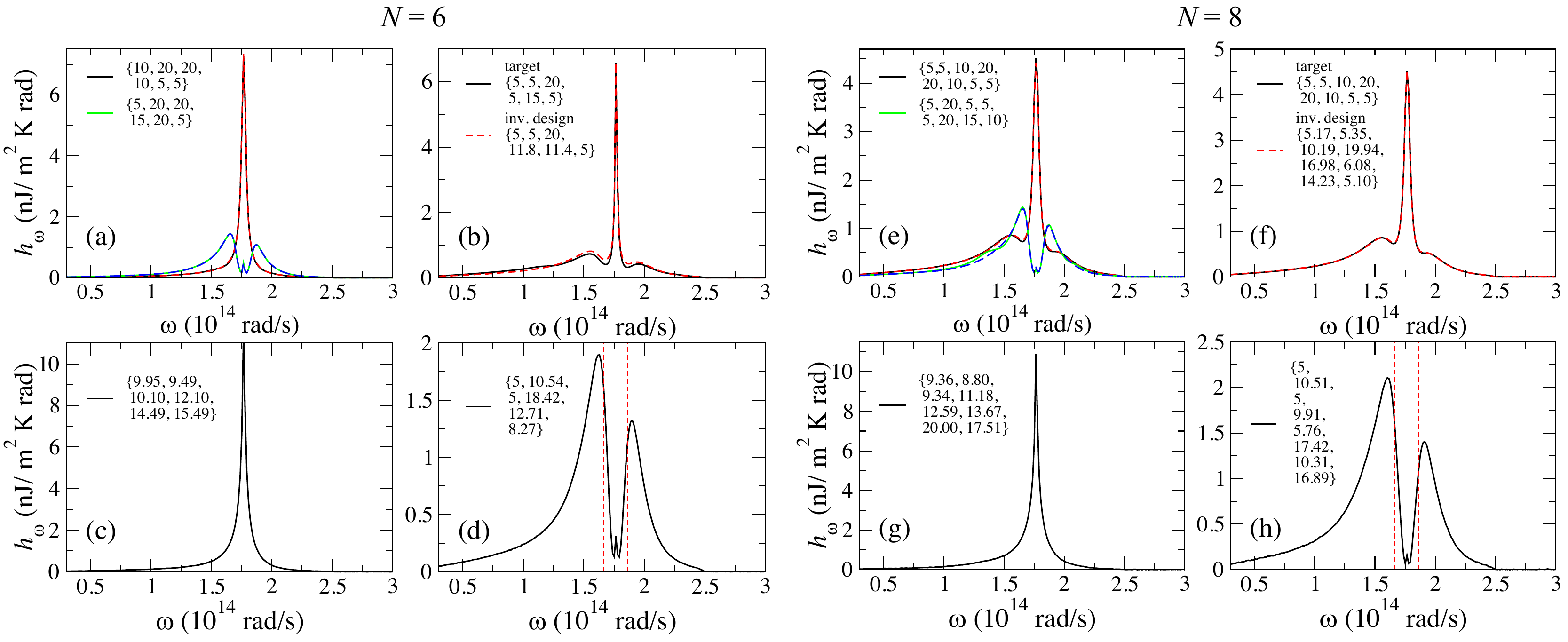}
\caption{(a-d) Same as in Fig.~\ref{fig-multilayer2} but now for $N=6$. (e-h) Same as in 
Fig.~\ref{fig-multilayer2} but for $N=8$.}
\label{fig-multilayer3}
\end{figure*}

We now want to illustrate how the NN can also be used to solve optimization problems. A first natural problem is 
to determine the layer structure (with $N=4$) that maximizes the total HTC $h$. Naively, in the limit $N \to \infty$
one expects to maximize the HTC by having all the layer thicknesses equal and equal to the gap size ($d_0 = 10$ nm)
\cite{Iizuka2018}, but for finite $N$ this is not necessarily the case and one cannot simply rely on physical intuition. 
This optimization problem can be easily solved by fixing the parameters of the NN, using the total HTC as the cost function 
to be maximized, and optimizing the network with respect to the input parameters (layer thicknesses). The result for this 
optimization problem is shown in Fig.~\ref{fig-multilayer2}(c) and the optimal thicknesses are $\{9.29, 9.78, 11.84, 14.43\}$ 
nm that lead to a total HTC at room temperature of $1.01 \times 10^5$ W/(m$^2$K). This value is $\sim 2.9$ larger than the 
HTC of the bulk system [$0.35 \times 10^5$ W/(m$^2$K)], which illustrates the fact that these multilayer systems can be used 
to further increase the NFRHT. Notice also that the optimal result is close to the case where all $d_i$ are equal to the
value of the gap size ($d_0$), which as mentioned above would the naive choice based on the idea of a quasi-periodic system,
which in turn would reduce the (Anderson-like) disorder in the structure.

Another interesting optimization problem consists in minimizing the heat transfer in a given frequency region, see 
Fig.~\ref{fig-multilayer2}(d), which might be motivated by the desire to inhibit the heat transfer in a certain frequency 
range. In this case, the cost function is defined as the ratio of the average of $h_\omega$ inside the range of interest 
and the corresponding average outside that region: $E = \bar h_{\omega,\mathrm{in}} / \bar h_{\omega,\mathrm{out}}$. The 
result of this optimization problem is shown in Fig.~\ref{fig-multilayer2}(d) and the corresponding layer thicknesses are 
$\{5.0, 12.44, 5.0, 19.32\}$ nm. Notice that the network suggests rather disparate thicknesses for the neighboring layers
and, in particular, the smallest possible thickness (within the range explored here) for the first metallic layer. This 
latter fact simply means that it is advantageous to have the first metallic layer as thin as possible to make it transparent
and avoid the contribution of the surface mode appearing in the layer 1-vacuum interface at $\omega_p/\sqrt{2} \approx 1.77 
\times 10^{14}$ rad/s, which corresponds to the center of the chosen gap.

We extended our analysis above to structures with 6 and 8 layers. As explained above, the presence of additional layers in 
the system makes possible to the appearance of extra surface states that may lead to the increase in the heat transfer
\cite{Iizuka2018}. On the other hand, from a deep learning standpoint, the additional degrees of freedom associated to 
the additional layers enables the used NNs to feature broader generalization abilities. For this purpose, we used the same 
network architecture (only changing the number of features in the input layer) and studied the same inverse and optimization 
problems. The results for these two structures are summarized in Fig.~\ref{fig-multilayer3}. The NNs were trained 
in these two cases with training sets containing 4825 $h_\omega$-spectra for $N=6$ and 65536 for $N=8$. Again, this 
amounts to an average of 4-5 thickness values per layer in the same range as for $N=4$. The training was done using the 
same cost function, optimizer, variable learning rate scheme, and the same division (80\%-20\%) between training and 
test set. The main difference in the training was the use of \emph{transfer learning}, a standard technique in deep 
learning which consists in using the parameters (weights and biases) of certain layers of a pre-trained network to train 
another network with similar architecture. In our case, we found that the best strategy is to use the parameters of all 
the layers of the network for $N=4$ ($N=6$) to initialize the parameters of the network for $N=6$ ($N=8$) and then 
proceed with the training. Since the networks for different values of $N$ have different number of neurons in the input 
layer, the extra parameters in the first layer were initialized randomly. With the help of transfer learning, we achieved 
average relative errors in the total integral of the $h_\omega$-spectra of 1.28\% (1.95\%) for the training set and 1.70\% 
(2.00\%) for the structure with $N=6$ ($N=8$). Without transfer learning the corresponding errors were 1.52\% (2.81\%) for 
the training set and 1.81\% (2.83\%) for the structure with $N=6$ ($N=8$), which is a significant difference. 

Another important point to highlight is that in the optimization problem, which consists in maximizing the total heat 
transfer resulted in an optimal structure for $N=6$ with thicknesses $\{9.95, 9.49, 10.10, 12.10, 14.49, 15.49\}$ nm with 
a HTC of $1.19 \times 10^5$ W/(m$^2$K), while for $N=8$ the optimal structure is $\{9.36, 8.80, 9.34, 11.18, 12.59, 
13.67, 20.00, 17.51\}$ nm with a HTC of $1.31 \times 10^5$ W/(m$^2$K). Notice that, as expected, the maximum HTC increases 
with the number of layers in the structure.

Let us conclude this Section by emphasizing that the study presented here can be extended to essentially any multilayer 
system, which may include anisotropic materials and metamaterials, in general, or the effect of external fields 
\cite{Moncada-Villa2015,Moncada-Villa2021}. Moreover, it could also be straightforwardly extended to deal with NFRHT 
between periodically patterned structures \cite{Fernandez-Hurtado2017} (this is exemplified in the following Section 
in the case of far-field emission).

\section{Passive radiative cooling} \label{sec-cooling}

Recently it has been shown that it is possible to cool down a device by simply exposing it to sunlight and 
without any electricity input~\cite{Rephaeli2013,Raman2014}. This striking phenomenon, referred to as \emph{passive 
radiative cooling}, is possible due to the fact that the Earth's atmosphere has a transparency window 
for electromagnetic radiation between 8 and 13 $\mu$m that coincides with the peak thermal radiation wavelengths 
at typical ambient temperatures. By exploiting this window one can cool a body on the Earth's surface by radiating 
its heat away into the cold outer space. While nighttime radiative cooling has been widely studied in the past,
the first proposal to realize this phenomenon during daytime was put forward in 2013 by Rephaeli 
\emph{et al.}~\cite{Rephaeli2013}. Inspired by nanophotonic concepts, these authors proposed a passive cooler that
consisted of two thermally emitting photonic crystal layers comprised of SiC and quartz, with a broadband 
solar reflector underneath. Subsequently, the same group designed and fabricated a multilayer photonic structure 
consisting of seven dielectric layers deposited on top of a silver mirror~\cite{Raman2014}. This design was shown to 
reach a temperature that is 5$^{\rm o}$C below the ambient air temperature, in spite of having about 900 W/m$^2$ 
of sunlight directly impinging upon it. After this first realization, there has been an intense research 
activity with the goal to optimize this daytime radiative cooling~\cite{Hossain2015,Gentle2015,Chen2016,
Kou2017,Zhai2017,Goldstein2017,Mandal2018,Li2019a,Zhou2019,Zhao2019,Li2019b,Li2021}. The goal of this Section is 
to illustrate how deep learning techniques can help in the theoretical design of devices for passive radiative
cooling.

Let us start by recalling the basics of the theory of passive radiative cooling following Refs.~\cite{Rephaeli2013,Raman2014}. 
Consider a radiative cooler of area $A$ at temperature $T$, whose spectral and angular emissivity is 
$\varepsilon(\lambda, \theta)$. When the radiative cooler is exposed to a daylight sky, it is subject to both 
solar irradiance and atmospheric thermal radiation (corresponding to an ambient air temperature $T_{\rm amb}$). 
The net cooling power $P_{\rm cool}$ of such a radiative cooler is given by the following power balance:
\begin{equation}
\label{eq-Pcool}
P_{\rm cool} = P_{\rm rad}(T) - P_{\rm atm}(T_{\rm amb}) - P_{\rm Sun} - P_{\rm cond+conv} .	
\end{equation}
Here, $P_{\rm rad}$ corresponds to the power radiated out by the structure and it
is given by
\begin{equation}
\label{eq-Prad}
P_{\rm rad}(T) = A \int d\Omega \cos \theta \int^{\infty}_0 d\lambda \, I_{\rm BB}(\lambda, T) 
\varepsilon(\lambda, \theta) ,
\end{equation}
where $\int d\Omega = 2\pi \int^{\pi/2}_0 d\theta \sin \theta$ is the angular integral over a hemisphere
and
\begin{equation}
I_{\rm BB} (\lambda, T) = \frac{2hc^2}{\lambda^5} \frac{1}{e^{hc/(\lambda k_{\rm B} T)} - 1} 	
\end{equation}
is Planck's distribution describing the spectral radiance of a blackbody at temperature $T$. On the 
other hand, $P_{\rm atm}$ is the absorbed power due to incident atmospheric thermal radiation and it 
is given by
\begin{multline}
\label{eq-Patm}
P_{\rm atm}(T_{\rm amb}) = A \int d\Omega \cos \theta \int^{\infty}_0 d\lambda \, I_{\rm BB}(\lambda, T_{\rm amb}) 
\times  \\ \varepsilon(\lambda, \theta)	 \varepsilon_{\rm atm}(\lambda, \theta)	,
\end{multline}
where $T_{\rm amb}$ is the ambient atmospheric temperature and $\varepsilon_{\rm atm}(\lambda, \theta)$ is the 
angle-dependent emissivity of the atmosphere \cite{Gemini}. The term $P_{\rm Sun}$ in Eq.~(\ref{eq-Pcool})
is the incident solar power absorbed by the structure and is given by
\begin{equation}
P_{\rm Sun} = A \cos (\theta_{\rm Sun}) \int^{\infty}_0 d\lambda \, I_{\rm AM1.5}(\lambda) 
\varepsilon(\lambda, \theta_{\rm Sun}) ,
\end{equation}
where $I_{\rm AM1.5}(\lambda)$ is the AM1.5 spectrum representing the solar illumination and $\theta_{\rm Sun}$
corresponds to the angle at which the structure is facing the Sun, which we shall assume to be zero. Finally, the term 
$P_{\rm cond+conv}$ in Eq.~(\ref{eq-Pcool}) is the power lost due to convection and conduction and adopts the form
\begin{equation}
\label{eq-P_non}
P_{\rm cond+conv}(T,T_{\rm amb}) = A \: h_c \: (T_{\rm amb} - T) , 
\end{equation}
where $h_c = h_{\rm cond} + h_{\rm conv}$ is a combined non-radiative heat coefficient that takes into account
the net effect of conductive and convective heating due to the contact of the cooler with external surfaces
and the air adjacent to the radiative cooler.

A given structure behaves effectively as a daytime cooling device when $P_{\rm cool} > 0$ at the ambient temperature, 
i.e., when the power radiated out by the cooler is greater than the combined effects of the incoming sources of heat 
from the Sun, atmosphere, and local conduction/convection. The power outflow $P_{\rm cool}(T=T_{\rm amb})$ then
defines the device's cooling power at ambient air temperature. Another important metric of the performance 
of the device is the equilibrium temperature, $T_{\rm eq}$, at which $P_{\rm cool} = 0$ in Eq.~(\ref{eq-Pcool}).
A radiative cooler with $T_{\rm eq}$ below the ambient temperature would cool an attached structure to a 
temperature below ambient over time.

\begin{figure}[t]
\includegraphics[width=\columnwidth,clip]{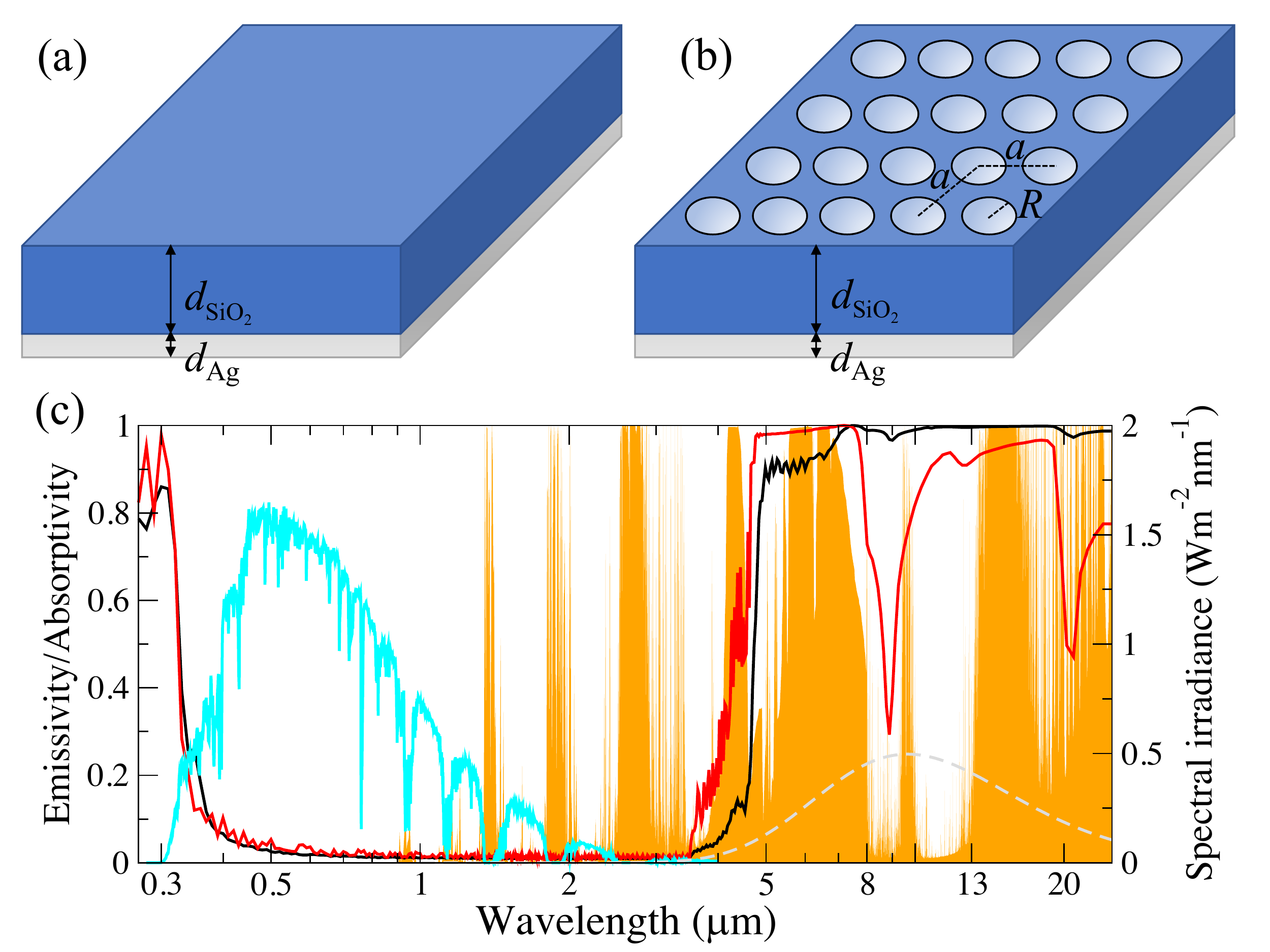}
\caption{(a) Schematics of the silica mirror used as a passive radiative cooler in Ref.~\cite{Kou2017}. It consists 
of a SiO$_2$ slab of thickness $d_\mathrm{SiO_2}$ and a silver thin film of thickness $d_\mathrm{Ag}$. (b) A 
nanostructured version of the cooler of panel (a), but featuring a periodic array of circular holes of radius 
$R$ with a lattice parameter $a$ (square lattice). (c) Emissivity as a function of the wavelength for a silica
photonic crystal (black solid line) and a silica mirror (red solid line) for $d_\mathrm{SiO_2} = 500$ $\mu$m and
$d_\mathrm{Ag} = 120$ nm. For the photonic crystal: $a = 100$ nm and $R = 50$ nm ($f = 0.785$). The cyan solid
line corresponds to the AM1.5 solar spectrum $I_{\rm AM1.5}$ (see right vertical axis), the orange curve to the 
atmospheric emissivity/absorptivity spectrum $\varepsilon_{\rm atm}$, and the gray dashed line to the blackbody 
radiation curve $I_{\rm BB}$ (50 times enlarged in spectral irradiance) at 300 K.}
\label{fig-PRC1}
\end{figure}

From the discussion above, it is obvious that the basic requirements for a structure to be a good passive cooler
are: \emph{(i)} to selectively emit thermal radiation in the atmospheric transparency window (from 8 to 13 $\mu$m), and,
\emph{(ii)} to reflect the solar radiation as much as possible. This requires to tune the emissivity over a very wide 
range (from the mid-infrared to the ultraviolet) and many complex structures have been proposed and realized. Here, 
we shall use as a starting point a rather simple configuration put forward in Ref.~\cite{Kou2017}, which consists of a 
silica mirror, see Fig.~\ref{fig-PRC1}(a), featuring a SiO$_2$ slab of thickness $d_\mathrm{SiO_2}$, which is responsible 
of a near-ideal blackbody in the mid-infrared, and a silver thin film of thickness $d_\mathrm{Ag}$, which provides 
reflection for the solar radiation. This simple structure with $d_\mathrm{SiO_2} = 500$ $\mu$m and 
$d_\mathrm{Ag} = 120$ nm was shown to achieve radiative cooling below ambient air temperature under direct sunlight 
($\sim 8$$^{\rm o}$C), which clearly outperforms more sophisticated designs \cite{Raman2014}. Furthermore, it was 
estimated that this cooler reached an average net cooling power of $\sim 110$ W/m$^2$ during daytime at ambient 
temperature even considering the significant influence of external conduction and convection \cite{Kou2017}. The 
device's performance was also shown to slightly improve by coating it with a polymer. In what follows, we shall 
investigate how the performance of this silica mirror can be improved via nanostructuration and how NNs can used to 
optimize its design.

\begin{figure}[t]
\includegraphics[width=\columnwidth,clip]{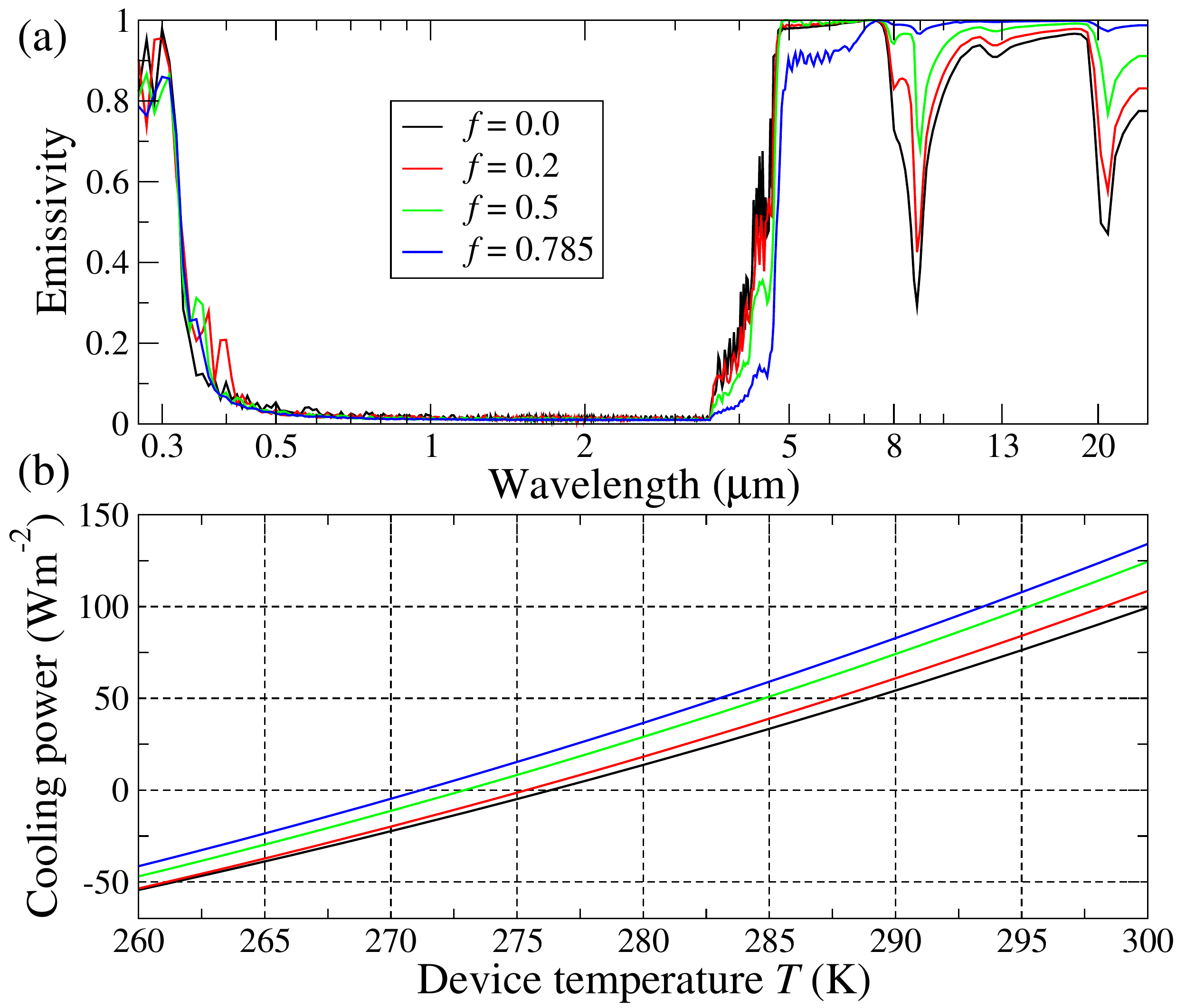}
\caption{(a) Emissivity of the photonic crystal cooling device as a function of the wavelength for several filling
factors $f$ and a hole radius $R = 50$ nm. The rest of the parameters are: $d_\mathrm{SiO_2} = 500$ $\mu$m and
$d_\mathrm{Ag} = 120$ nm. (b) The corresponding cooling power density as a function of the device temperature under 
AM1.5 illumination. The ambient temperature of the atmosphere is taken to be $T_\mathrm{amb} = 300$ K.}
\label{fig-PRC2}
\end{figure}

To be precise, we shall now discuss how the introduction of a periodic array of air circular holes in the silica
layer can boost the performance of the device as a passive cooler, see Fig.~\ref{fig-PRC1}(b). We consider 
the case of a square lattice with a lattice parameter $a$ and holes of radius $R$. We define the filling factor $f$
as the fraction of the area occupied by the air holes: $f = \pi (R/a)^2$, which varies between 0 (no holes) and 
$\pi/4 \approx 0.785$ (for the largest possible holes, $R=a/2$). The basic idea is that this photonic crystal
can enhance the emissivity of silica in the atmospheric transparency window, while maintaining the sunlight 
absorption of the unstructured silica mirror \cite{Zhu2015}. The infrared enhancement of the photonic crystal is 
simply due to a reduction of the impedance mismatch between the silica surface and the surrounding air. This effect 
is illustrated in Fig.~\ref{fig-PRC1}(c) where we compare the normal-incidence emissivity of a silica photonic crystal 
with the corresponding silica mirror with no holes. For reference, we also show in that figure the AM1.5 spectrum, 
the normal-incidence of the atmosphere's emissivity and Planck's blackbody distribution. As one can see, the photonic 
crystal device behaves almost as a black body in the relevant infrared region, which is precisely what we are looking
for. 

We illustrate in more detail the role of the nanostructuration in Fig.~\ref{fig-PRC2}(a) where we show how the
emissivity of a photonic crystal device progressively increases in the atmospheric transparency window upon increasing
the filling factor. In this example we set $R = 50$ nm, $d_\mathrm{SiO_2} = 500$ $ \mu$m, and an ambient temperature 
$T_\mathrm{amb} = 300$ K. The emissivity increase in the mid-infrared results in an improvement of the performance of 
the cooling device as shown in Fig.~\ref{fig-PRC2}(b) where we display the corresponding cooling power $P_\mathrm{cool}$ 
as a function of the device temperature $T$. Here, we have ignored the contribution of non-radiative processes (conduction 
and convection) and approximated the emissivities by the normal incidence results (this approximation will be used 
throughout this Section). Notice that upon increasing the filling factor, the cooling power can increase up to a 40\% 
at ambient temperature (300 K) and the equilibrium temperature $T_\mathrm{eq}$ can be reduced by $\sim 5$$^{\rm o}$C, 
as compared to the unstructured silica mirror. 

All the results in this work for the emissivity of the periodically patterned structures were calculated using
the rigorous coupled wave analysis (RCWA) described in Ref.~\cite{Caballero2012}. It is important to stress that this
is a numerically exact method that makes use of the so-called fast Fourier factorization when dealing with the Fourier 
transform of two discontinuous functions in the Maxwell equations. This factorization solves the known convergence 
problems of the RCWA approach, see Ref.~\cite{Caballero2012} for details, and its use was critical in this case because
we deal with materials with very different dielectric functions and the emissivities have to be computed over a huge 
wavelength range (from the UV to the mid-infrared). Owing to its ability to generate training sets in a robust
and efficient way, our own implementation of the RCWA method became a key ingredient in the successful application 
of deep techniques in this context. 

Finally, we point out that for all the parameter values considered here, we made sure that the emissivity spectra were 
converged up to a 1\% relative error for every wavelength point. This required to take into account up to several 
thousand plane waves for the shortest wavelength (UV/visible range) and the largest holes. To give an idea of the 
required computational time, the most time-consuming emissivity spectra took about 24 hours in a desktop computer 
with a 2.3-GHz Intel Xenon processor running in parallel in 18 CPUs. For the RCWA calculations we used as an input 
the dielectric function of SiO$_2$ tabulated in Ref.~\cite{Palik1985} and for Ag that of Ref.~\cite{Yang2015}.

To study systematically the role of the nanostructuration in the performance of the cooler, we define the following 
optimization problem. We consider as input parameters the silica layer thickness $d_\mathrm{SiO_2}$, the filling
factor $f$, and the hole radius $R$ (we fix the thickness of the back reflector to $d_\mathrm{Ag} = 120$ nm), and we 
search for the optimal values of these parameters to maximize the cooling power at ambient temperature. 
To solve this problem with the help of NNs, we first constructed a training set
with RCWA calculations of the emissivity of the cooler, which were then used to compute $P_\mathrm{cool}$ using 
Eqs.~(\ref{eq-Pcool})-(\ref{eq-P_non}). The training set contained 900 emissivity spectra
with 15 different values of $d_\mathrm{SiO_2}$ between 1 and 2000 $\mu$m, 10 different values of $f$ between 0 and
$\pi/4$, and 6 values of $d_\mathrm{SiO_2}$ from 30 to 200 nm. Every spectrum contains the emissivity for 525
wavelength values ranging from 270 nm to 25 $\mu$m. A preliminary analysis using this training set indicated that 
below hole radii of 100 nm the results are fairly insensitive to the exact radius value, while for larger holes the 
sunlight absorption increases and the performance of the device decreases drastically. For this 
reason, we fixed the radius value to $R = 30$ nm, and reduced the input parameters to $d_\mathrm{SiO_2}$ and $f$ 
(the filling factor). Thus, our training set contained in practice 150 emissivity spectra, which were divided into 
an actual training set (80\%) and a test set (20\%).

\begin{figure}[t]
\includegraphics[width=\columnwidth,clip]{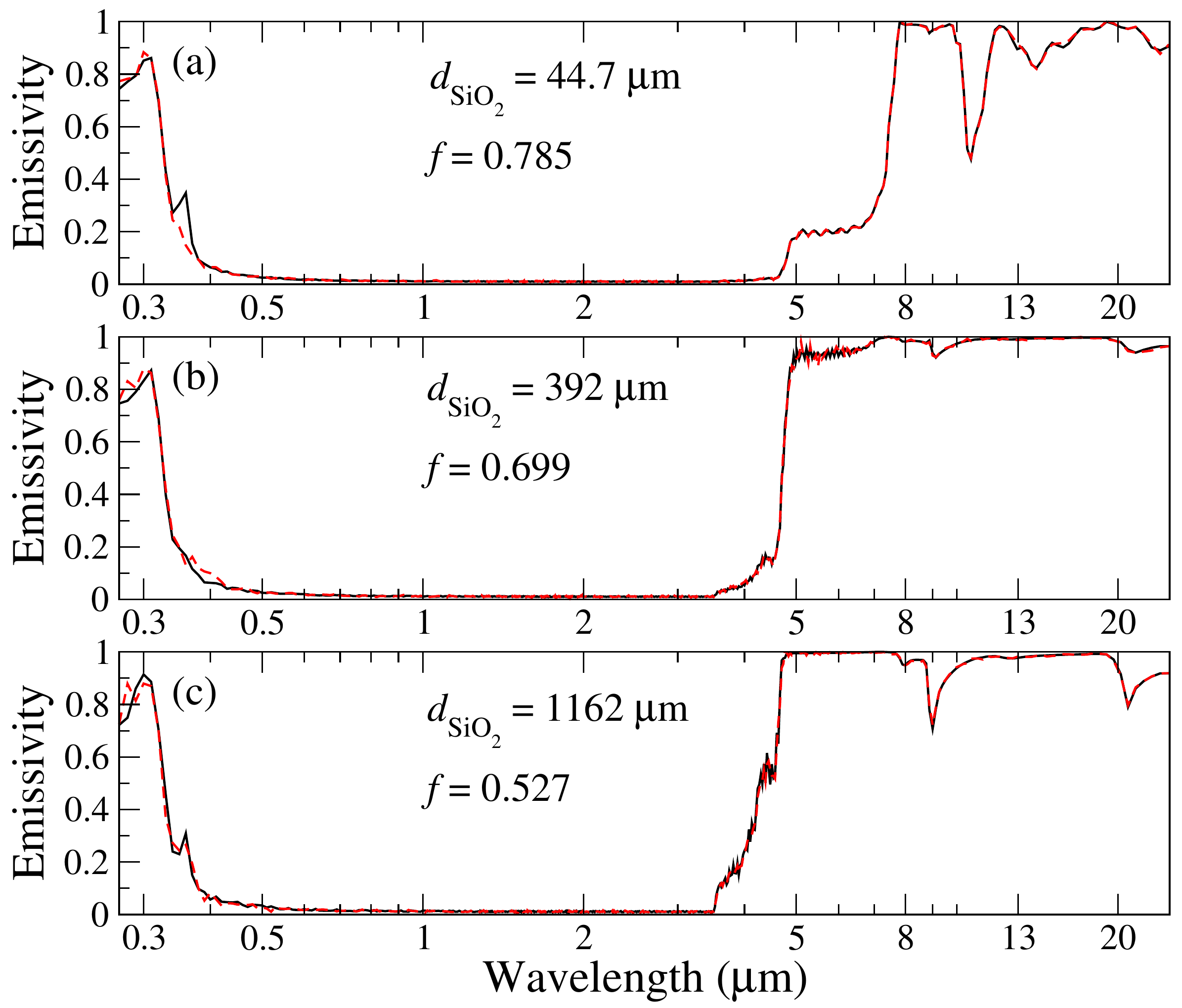}
\caption{(a-c) Comparison between the emissivity spectra of silica photonic crystal devices computed with the RCWA 
method (solid lines) and the prediction of the NN (dashed lines) for different values of the silica layer thickness 
and filling factor, as indicated in the legends. The hole radius is $R = 30$ nm and $d_\mathrm{Ag} = 120$ nm.}
\label{fig-PRC3}
\end{figure}

For this problem we found that a NN with 3 hidden layers (with 250 neurons per layer) is enough to satisfactorily 
reproduce the training set. This network contains two neurons in the input layer (corresponding to the two input 
parameters or features in this problem), while the output layer has 525 neurons corresponding to the wavelength values 
in the emissivity spectra. The NN was trained during 50,000 epochs using the MSE as the cost function, 
the ADAM optimizer, and the ReLU activation function in all layers, except in the output one. No early stopping was 
used in this case. After training, the MSE for the training set was $4.6 \times 10^{-5}$ and $4.4 
\times 10^{-4}$ for the test set. In Fig.~\ref{fig-PRC3} we illustrate the ability of this NN to reproduce emissivity 
spectra of the test set. As seen, the NN is able to accurately reproduce very different spectra over the entire 
wavelength range, which demonstrates the ability of our NN to generalize to cases it was not trained on. It is also 
remarkable to find that degree of accuracy despite the moderate size of our training set (as mentioned,
formed by just 150 samples).

\begin{figure}[t]
\includegraphics[width=\columnwidth,clip]{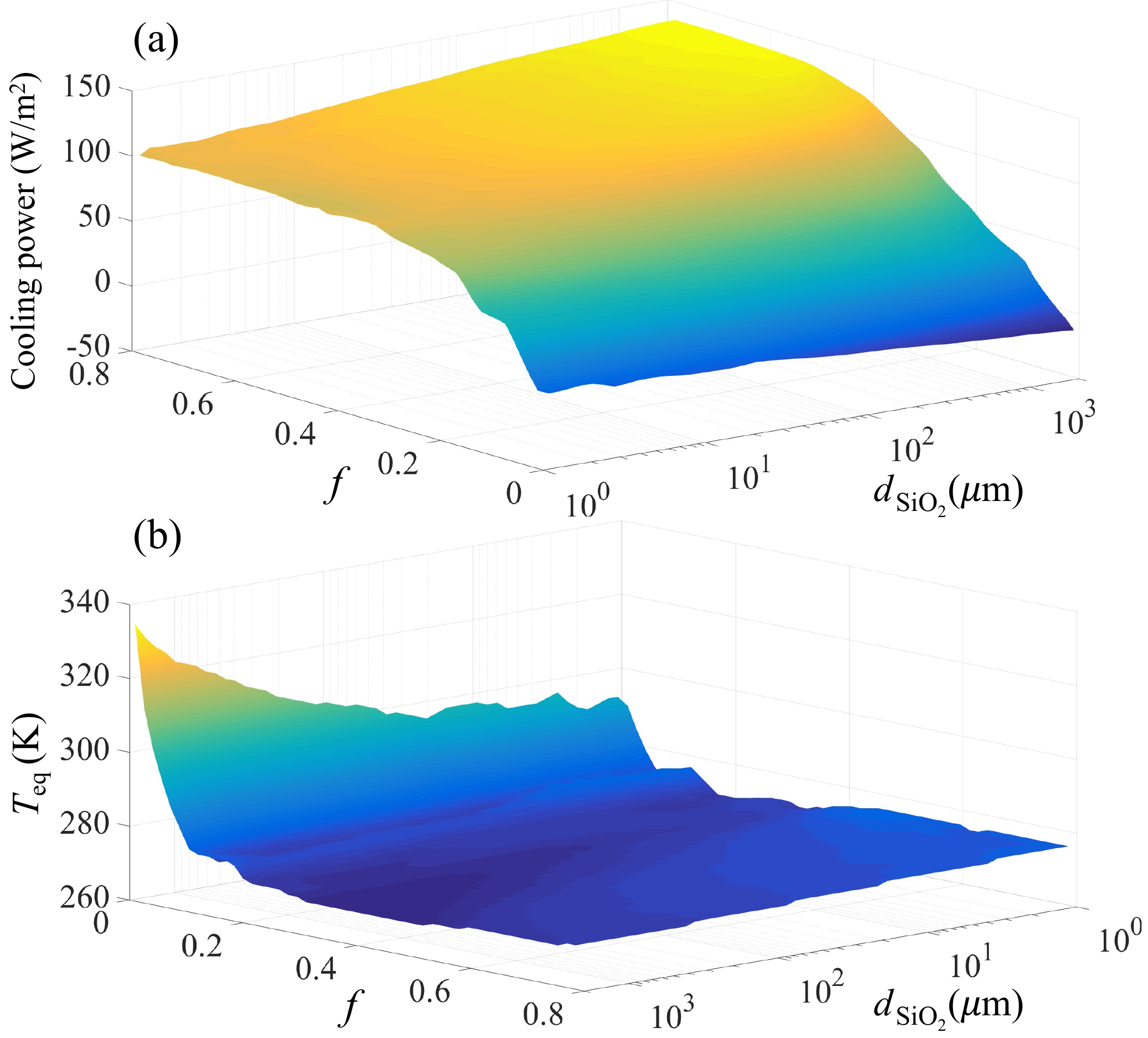}
\caption{(a) Cooling power density as a function of the silica thickness layer and filling factor of photonic 
crystal cooling devices as computed with the trained NN. In this case, the device temperature and the ambient
temperature were assumed to be 300 K. The contribution of non-radiative processes was neglected. (b) The
corresponding equilibrium temperature.}
\label{fig-PRC4}
\end{figure}

Next, we used the NN as a computational engine to solve our optimization problem, namely the maximization of the 
device's cooling power. As explained in the previous Section, this optimization process is very efficient because 
we can analytically compute the NN gradients with respect to the inputs using backpropagation \cite{Peurifoy2018}. 
In Fig.~\ref{fig-PRC4}(a) we illustrate the results predicted by the NN for the cooling power density as a function 
of the filling factor and the silica layer thickness for $T = T_\mathrm{amb} = 300$ K. For these calculations, and 
since we are mainly interested in comparing with the results for the unstructured silica mirror, we ignored the 
contribution of the non-radiative processes (conduction and convection). For completeness, we also show in 
Fig.~\ref{fig-PRC4}(b) the corresponding equilibrium temperature. Let us remind that, contrary to the
cooling power, the equilibrium temperature is very sensitive to the contribution of conduction and convection.
As it is obvious from Fig.~\ref{fig-PRC4}(a), the optimization process suggests that the optimal parameters are
$f = 0.785$ and $d_\mathrm{SiO_2} = 2$ mm. This means that the filling factor has to be as large as possible, which
confirms the naive expectation. With respect to the silica layer thickness, there is no significant difference in the 
cooling power in the range $d_\mathrm{SiO_2} \in [500\, \mu\mbox{m}, 2\, \mbox{mm}]$.

The study presented in this Section can be generalized in a number of ways. For instance, we also analyzed the role of 
a finite depth of the air holes (in the calculations discussed above, it was assumed that the silica layer was perforated
all the way down to the Ag thin film). In particular, we found that for the optimal parameters found above, a hole depth of 
around 10 $\mu$m is enough to reach values for the cooling power very similar to those obtained for a fully perforated 
silica layer. Of course, there are a number of different Bravais lattices that one could also explore, as well as many 
different shapes of the air holes. In the case of regular hole shapes one could still use sequential fully connected 
NNs like in this work, but if one wants to consider arbitrary shapes it might be more convenient to use two-dimensional 
convolutional neural networks (CNNs) \cite{Gu2018}, which have already been successfully used in the context of 
nanophotonics for the design of metasurfaces with desired properties 
\cite{So2020,Hegde2020,Jiang2020,Ma2021,Piccinotti2021,Liu2021}. 

\section{Thermal emission of subwavelength objects} \label{sec-emission}

The goal of this Section is to show how NNs can also be helpful in the context of the description of the
thermal emission of a single object of arbitrary size and shape. In particular, we are interested in 
the thermal emission of subwavelength objects, i.e., objects in which some of their dimensions are
smaller than the thermal wavelength $\lambda_{\rm Th}$. Part of the interest in this problem lies in the 
fact that, as already acknowledged by Planck in his seminal work~\cite{Planck1914}, Planck's law fails to describe 
the thermal properties of subwavelength objects simply because it is based on ray optics. In this sense, one may 
wonder whether the blackbody limit for the thermal emission of a body (given by Stefan-Boltzmann's law) can be 
overcome in the case of subwavelength objects, something that is not possible in the case of infinite objects. 
Actually, it is well-known that the emissivity of a finite object can be greater than 1 at certain frequencies 
\cite{Bohren1998,Schuller2009}, but that is not enough to emit more than a black body. In fact, only a modest 
super-Planckian thermal emission has been predicted in rather academic situations \cite{Kattawar1970,Golyk2012}, 
and it has never been observed. Recently, Fern\'andez-Hurtado \emph{et al.}~\cite{Fernandez-Hurtado2018} 
showed that elongated objects with subwavelength dimensions can indeed have directional emissivities much larger
than 1, which can lead to super-Planckian far-field radiative heat transfer between two of those bodies, as
it has been experimentally verified \cite{Thompson2018}. However, the total thermal emission of those objects is 
still smaller than that of a black body. There are by now several experiments showing the failure of Planck's law in 
the description of the thermal emission of subwavelength objects (although no super-Planckian emission has yet been 
observed). For instance, Ref.~\cite{Wuttke2013} reported this failure in the case of small optical fibers, while 
Ref.~\cite{Shin2019} did it in the case of nanoribbons made of silica with a thickness of 100 nm, much 
smaller than both $\lambda_{\rm Th}$ and the skin depth, while the other dimensions could be much larger. From the 
theory side, the description of the thermal emission of a single object of arbitrary size and shape continues to 
be very challenging and there is only a handful of general-purpose numerical approaches that can tackle this problem 
\cite{Otey2014,Rodriguez2013,Reid2015,Polimeridis2015,Martin2017}. These techniques are often exceedingly time 
consuming and systems like the nanoribbons explored in Ref.~\cite{Shin2019} are still out of the scope of these 
techniques. Thus, in the rest of this Section we shall show how the use of NNs can contribute to alleviate this 
situation.

The total power emitted by any object at a temperature $T$ is given by \cite{Modest2013}
\begin{equation}
\label{eq-Pem}
P_{\rm em} = \pi A \int^{\infty}_0 d\omega \, I_{\rm BB}(\omega,T) \varepsilon(\omega)	 ,
\end{equation}
where $A$ is the total area of the object, $\varepsilon(\omega)$ is the angular-averaged frequency-dependent 
emissivity of the body, and $I_{\rm BB}(\omega,T)$ is the frequency-dependent Planck distribution given by 
\begin{equation}
I_{\rm BB}(\omega,T) = \frac{\omega^2}{4\pi^3 c^2} \frac{\hbar \omega}{e^{\hbar \omega/ k_{\rm B}T} -1} .
\end{equation}
In the case of a black body, $\varepsilon(\omega) = 1$ for all frequencies and the total emitted power is 
given by Stefan-Boltzmann law: $P_{\rm em, BB} = A \sigma T^4$, where $\sigma = 5.67 \times 10^{-8}$ W/(m$^2$K$^4$).

\begin{figure}[t]
\includegraphics[width=\columnwidth,clip]{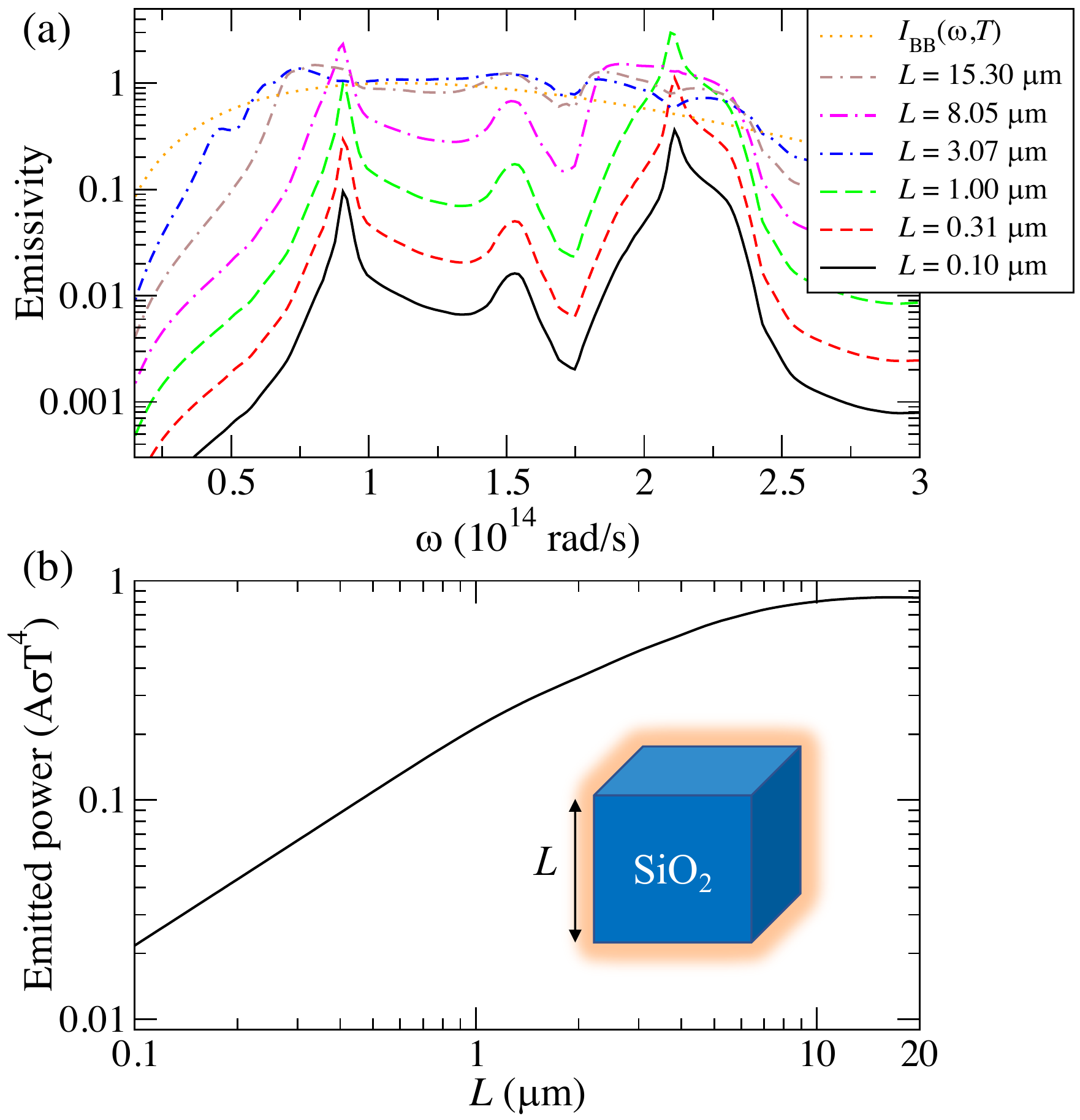}
\caption{(a) Emissivity of a silica cube of side $L$ as a function of the frequency. The dotted orange line corresponds
to Planck's distribution $I_{\rm BB}(\omega,T)$ at $T= 300$ K in arbitrary units. (b) Total power emitted by a SiO$_2$ 
cube (see inset) as a function of its side $L$ at $T = 300$ K. The power is normalized with the blackbody result, 
$A \sigma T^4$, where $A=6L^2$ is the total area of the cube.}
\label{fig-cubes1}
\end{figure}

To illustrate the use of NNs in this particular context, we consider here a proof-of-principle example, namely 
the thermal emission of a silica cube of arbitrary side $L$ and at room temperature ($T=300$ K), 
see inset of Fig.~\ref{fig-cubes1}(b). A cube is already a sufficiently complicated geometry such that its 
emissivity cannot be calculated analytically. We have computed this emissivity using the numerical approach
known as thermal discrete dipole approximation (TDDA), as described in section IV of Ref.~\cite{Martin2017}.
In this approach, an object is discretized in terms of point dipoles in the spirit of the DDA method that is 
widely used for describing the scattering and absorption of light by small particles \cite{Purcell1973,Yurkin2007}.
In our case, we modeled silica cubes with sides ranging from 0.1 $\mu$m (much smaller than the thermal wavelength) 
to 20 $\mu$m, which is comparable to the thermal wavelength. We discretized the cubes in terms of a lattice of cubic 
dipoles and used up to $\sim 12000$ dipoles, which was checked to be enough to accurately converge the results even 
for the largest cubes considered here. The calculation of a single spectrum with 200 frequency
points and with a discretization with $\sim 12000$ dipoles took about 24 hours in a desktop computer with a 2.3-GHz 
Intel Xenon processor running in parallel in 18 CPUs. The dielectric function of SiO$_2$ was taken from the tabulated 
values in Ref.~\cite{Palik1985}. 

A representative set of examples of the computed emissivity are shown in Fig.~\ref{fig-cubes1}(a) for 
various values of $L$. Notice that for certain frequencies the emissivity can be larger than 1 (e.g., for $L=1\, \mu$m 
and $\omega \sim 2.1 \times10^{14}$ rad/s, $\varepsilon \sim 2.9$). However, this does not mean that a silica cube 
can be a super-Planckian emitter. As we show in Fig.~\ref{fig-cubes1}(b) where one can see the total emitted power as 
a function of $L$, a cube always emits less than a black-body cube of the same size. Notice also that for small 
cubes the emitted power is proportional to the cube volume, which is due to the fact that in this regime the silica 
skin depth at the relevant frequencies is larger than $L$ \cite{Krueger2011,Krueger2012}. This means that the whole 
object contributes to the thermal emission. However, as the size increases, the emitted power becomes proportional 
the cube area and it tends to converge to the value of an infinite silica surface, which with the optical constants 
used here is equal to 0.79 at room temperature. This behavior reflects the fact that when $L$ becomes larger than 
the skin depth, the thermal emission only originates from the surface, as it happens in macroscopic objects.

Now we show that a NN can learn the emissivity spectra of a cube. For this purpose, we have computed
a training set of 100 emissivity spectra with $L \in [0.1 \, \mu\mbox{m}, 20 \, \mu\mbox{m}]$ with equally spaced 
side values in a logarithmic scale. Additionally, we have calculated another 20 spectra in the same range to form 
the test set. Later on we shall explore what happens when varying the size of the training set. In this case, we 
have done a hyperparameter search (changing the number of layers, the number of neurons per layer, the learning rate, 
etc.), and used $k$-fold cross-validation (with $k=5$) to select the optimal hyperparameters \cite{James2017}. 
The only input feature in this was the cube side (i.e., we have a single neuron in the input layer), and the output was 
the emissivity spectrum sampled at 200 equidistant points between $0.15 \times 10^{14}$ and $3.04 \times 10^{14}$ rad/s 
(i.e., the output layer has 200 neurons). The NNs were trained using the MSE as the cost function, the ADAM optimizer, 
and the ReLU activation function in all layers, but the output one where no activation function was used as it is 
customary in a regression problem. The NNs were trained for a maximum of $\sim50,000$ epochs and we used early 
stopping based on the validation error to conclude the training. We found that an optimal NN is composed of 
4 hidden layers with 250 neurons per layer, which is the network we used for all the calculations that we are
about to describe. 

\begin{figure}[t]
\includegraphics[width=\columnwidth,clip]{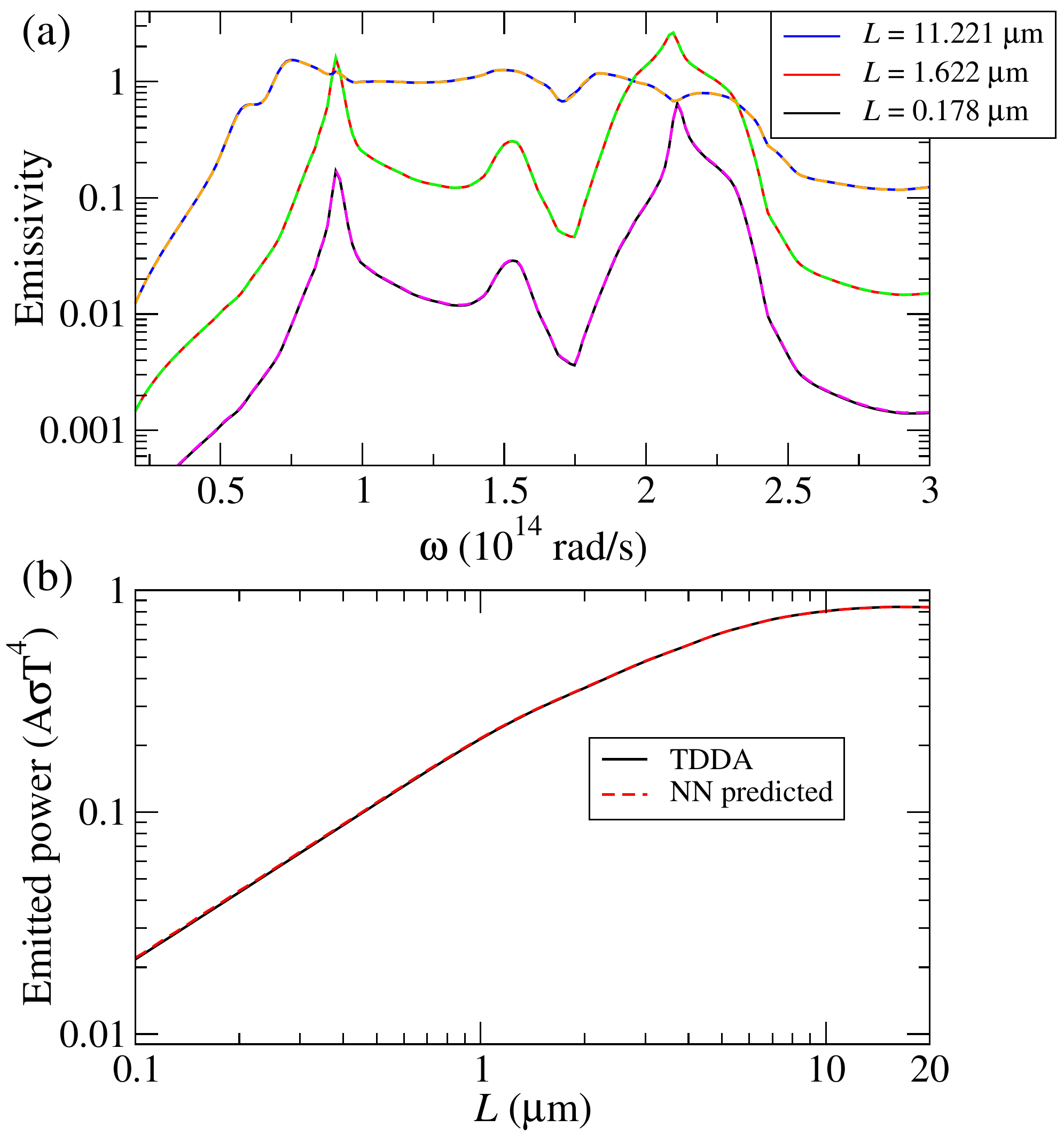}
\caption{(a) Comparison between the real emissivity spectra of a SiO$_2$ cube computed with TDDA (solid lines) and 
the prediction of the NN (dashed lines) for various values of cube side $L$. (b) The corresponding comparison for the 
total power emitted by a SiO$_2$ cube as a function of its side $L$ at $T = 300$ K. The power is normalized with the 
blackbody result.} 
\label{fig-cubes2}
\end{figure}

As in the previous examples, the first application was to test the forward computation of the network to see how well 
it reproduces the emissivity spectra it was not trained on. This is illustrated in Fig.~\ref{fig-cubes2}(a) where we 
show that the NN can very accurately reproduce several representative examples of the test set. We also show in 
Fig.~\ref{fig-cubes2}(b) that using the NN predictions for the emissivity spectra and Eq.~(\ref{eq-Pem}), we can 
accurately reproduce the size-dependence of the total emitted power. To be more quantitative, we computed the
average relative error per point in the emissivity spectra and found that is equal to $0.65$\% for the training
set and $0.63$\% for the test set, which demonstrates the excellent generalization ability of the optimal NN. 

\begin{figure}[t]
\includegraphics[width=\columnwidth,clip]{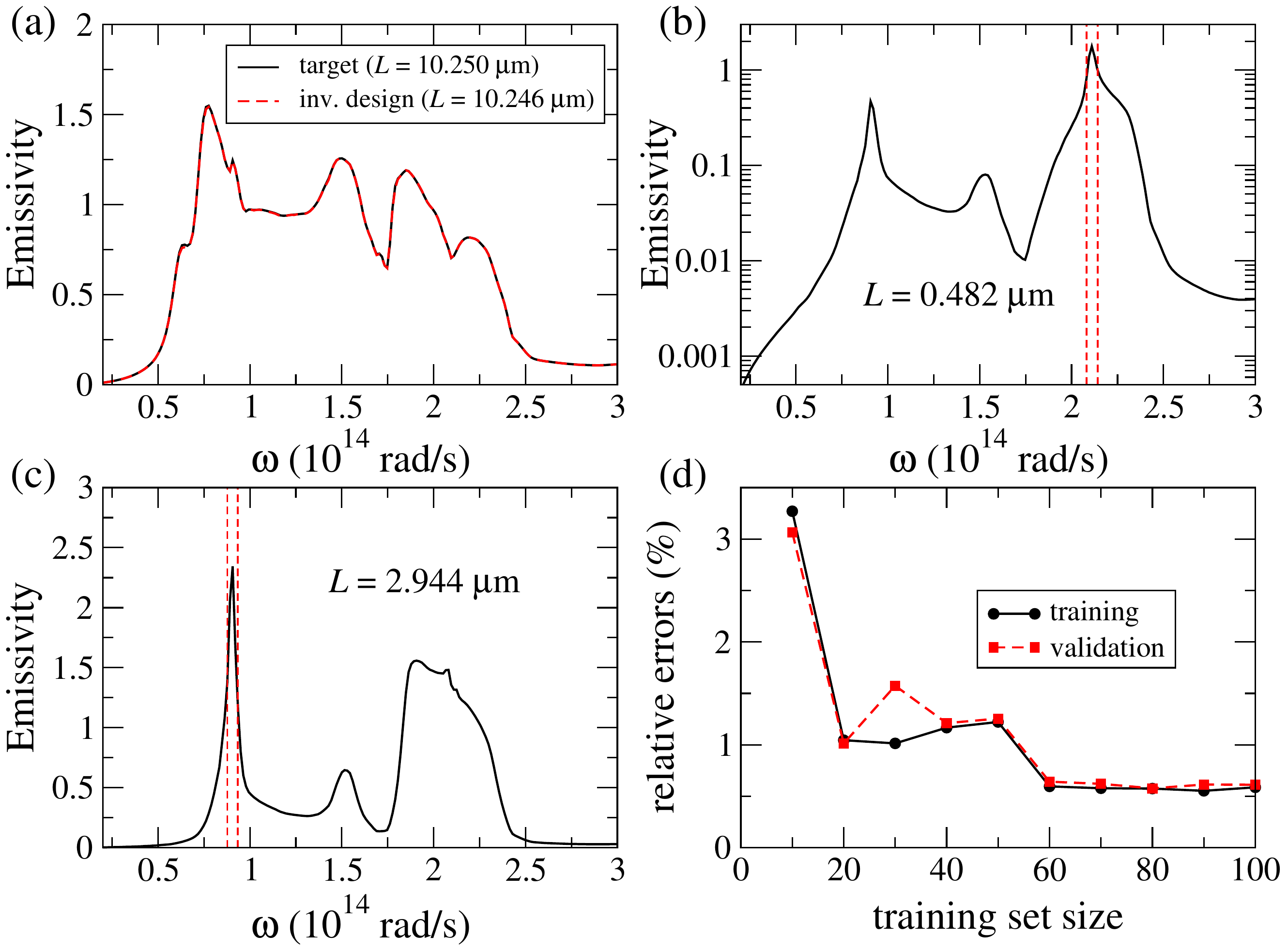}
\caption{(a) Comparison of the NN approximation for the emissivity of a SiO$_2$ cube to a target spectrum for 
$L=10.250$ $\mu$m following the inverse design problem described in the text. (b,c) Result of the optimization
problem where the emissivity of a SiO$_2$ cube in the frequency range defined by the vertical dashed lines is
maximized while the emissivity outside is minimized. The obtained value of the optimal cube side is indicated in 
the panels. (d) Learning curves showing the mean relative error (training and validation) as a function of the 
training set size for the optimal NN simulating the emissivity of a SiO$_2$ cube.} 
\label{fig-cubes3}
\end{figure}

Now, we illustrate the possibility to use the NN to do inverse design. The goal is to show that with the help of the NN we 
can find the geometry (the value of the cube side) that would be able to reproduce an arbitrary emissivity spectrum. Again, 
the idea is to keep fixed all the parameters of the NN and use backpropagation to train the inputs. This is done by fixing 
the output to the desired output and iterating the input to minimize the difference between the spectrum predicted by 
the NN and the target spectrum (i.e., the cost function in this case is simply defined as the MSE between the predicted 
and the target spectrum). As also pointed in the previous Sections, this minimization process is extremely efficient 
because we can analytically compute the NN gradients of the cost function with respect to the inputs using 
backpropagation \cite{Peurifoy2018}. After converging this process, the NN suggests a geometry to reproduce the target 
spectrum. The inverse design ability of our NN is illustrated in Fig.~\ref{fig-cubes3}(a) where the target spectrum 
was randomly chosen to be that of a cube of $L=10.250$ $\mu$m (to ensure that we have a physically realizable spectrum). 
As observed, the NN is able to accurately reproduce this target spectrum and suggests that the corresponding cube side 
is $L=10.246$ $\mu$m, which is in excellent agreement with the actual value. We have obtained similar results with 
all the target spectra explored in the range of cube sides used to train the network.  

Next, we illustrate the fact that the NN can also be used to solve optimization problems. In particular, we aim at determining
what is the optimal cube side to maximize the emissivity at a given narrow frequency range while minimizing the emissivity
outside this range. For this purpose, we fix the parameters of the NN, define a cost function for this task, and 
optimize the network with respect to the input parameters (the cube side in this problem). A convenient cost function in this 
case is defined as the ratio of the average of the emissivity inside the range of interest and the corresponding average 
outside that region: $E = \bar \varepsilon_\mathrm{in} / \bar \varepsilon_\mathrm{out}$. The results for this optimization
problem for two ranges around the frequencies of the silica phonon polaritons are shown in Fig.~\ref{fig-cubes3}(b,c),
where we also indicate the corresponding optimal values of the cube side $L$. Notice that for the high-frequency 
phonon polariton (panel b), the emission is maximized for a relatively small cube, in which the emission comes from 
the whole body. For the low-frequency resonance (panel c) such an emission is maximized for a cube of size comparable 
to the skin depth, in which thermal emission is mainly a surface phenomenon.

The use of neural networks becomes particularly useful when there is lack of real (or training) data. In this sense, one
may wonder how large the training set has to be in this case for the NN to be able to generalize well, i.e., to accurately 
predict spectra that have not been used in the training procedure. To answer this question, we analyzed the performance
of our optimal NN as a function of the training set size. The corresponding learning curves are shown in 
Fig.~\ref{fig-cubes3}(d) where one can see both the training and validation error expressed as the mean percent off per 
point on the spectrum. As seen, even training the NN with as little as 20 spectra, the validation error is on the order 
of $1$\%, which illustrates how efficient NNs are at learning complex patterns like our emissivity spectra. Notice also 
that the evolution of the training and validation errors, which are very similar, shows that there is little overfitting, 
irrespective of the training set size. 

Finally, having shown that simple NNs can efficiently learn the emissivity spectra of small objects, we also explored
how to extend their use in this context. Thus, for instance, we tried and succeeded in using the NN 
trained on silica cubes to study the thermal emission of other simple objects such as spheres. Following the idea of transfer 
learning, we used the weights and biases of our NN as an initial starting point to train a network with the same architecture 
to learn the emissivity spectra of silica spheres of arbitrary radius (not shown here). As expected, this type of transfer 
learning substantially improves the accuracy of the training process and provides a promising path to model more demanding
structures. In this sense, we are currently investigating if such an approach may be allow us to model structures
like the silica nanoribbons mentioned above \cite{Shin2019}, which still remain out of the scope of any current theoretical 
method. 

\section{Conclusions} \label{sec-conclusions}

In summary, in this work we have reported the, to our knowledge, first systematic study of the application of deep learning 
techniques to the theoretical analysis of different radiative heat transfer phenomena. In particular, we have applied deep 
artificial neural networks to three state-of-the-art problems, ranging from near-field radiative 
heat transfer between multilayer systems to the description of the far-field thermal emission of extended systems in the context 
of passive radiative cooling and finite systems of arbitrary shape that defy Planck's law. Despite the significant differences 
of the three studied scenarios, in all of them we have shown that, after training them on datasets of moderate size, simple 
neural network architectures can be used to do fast simulations of a great variety of thermal processes with a high precision. 
Moreover, we have demonstrated that neural networks can also be used as computational engines to solve interesting inverse 
design and optimization problems in the context of radiative heat transfer. 

In this work, we have focused on proof-of-principle examples with the goal to illustrate some of the main ideas. We believe 
the concepts put forward here can be generalized to deal with much more complex structures and phenomena. Thus, for
instance, it would of great interest to use the ideas discussed here in the context of heat transfer in many-body systems, 
a vast topic that is currently attracting a lot of attention \cite{Biehs2021}. Although we have mainly focused on 
the use of neural networks as function approximators, this is by no means the only possibility. It would 
also be very interesting to apply generative models to thermal radiation problems based on techniques like variational 
autoencoders (VAEs) or generative adversarial networks (GANs) \cite{Goodfellow2016}. Those techniques could, for instance, 
help to generate new data in situations where it is very hard to provide extensive training sets. This, in turn, could 
help to better train neural networks via data augmentation. On the other hand, less conventional networks like 
recurrent neural networks (RNNs) might find applications in the modeling of time-dependent thermal phenomena or in the
development of protocols for thermal management. Overall, we believe that the application of deep learning to radiative 
heat transfer is still in its infancy and we hope this work can stimulate further research work aimed at exploring 
how artificial neural networks, and more generally artificial intelligence techniques, can contribute to accelerate the 
advance of this field.

\acknowledgments

J.J.G.E. was supported by the Spanish Ministry of Science and Innovation through a FPU grant (FPU19/05281).
J.B.A.\ acknowledges financial support from Ministerio de Ciencia, Innovaci\'on y Universidades (RTI2018-098452-B-I00).
J.C.C.\ acknowledges funding from the Spanish Ministry of Science and Innovation (PID2020-114880GB-I00).

\end{document}